\documentclass[letterpaper,12pt]{article}
\usepackage{fullpage}
\usepackage{amsmath, amsthm, amssymb}
\usepackage{graphicx}
\usepackage{verbatim}
\usepackage{fancyhdr}
\usepackage[colorlinks=true,linkcolor=blue]{hyperref}
\usepackage[all]{xy}

\theoremstyle{plain}
\theoremstyle{definition}

\graphicspath{{Images/}}


\title{\textbf{Predicting Block Halving Party Times}}

\author{Meni Rosenfeld\footnote{The author can be contacted at meni@bitcoin.org.il. If you would like to support this research, you can send bitcoins to 18Vftuz74RnVdMM1boN69oW59FagBNSvo4.}\\
}

\date{June 3, 2016\\
Latest version: \today}

\begin{document}

\maketitle
\begin{abstract}
Bitcoin (\cite{Bitcoin}) is the world's first decentralized digital currency. The rate at which bitcoins enter circulation is cut in half every 4 years, approximately. These events are considered landmarks in Bitcoin's history, and as such are widely celebrated. However, this requires placing confidence intervals on the precise timing of the halving well in advance, and the particular mechanism by which the halving time is determined makes this challenging. In this paper, we intend to help party planners by describing the problem, and highlighting several techniques to estimate the mean and variance of the halving.
\end{abstract}
\pagenumbering{roman}
\tableofcontents
\section{Introduction}\label{chap:intro}
\pagenumbering{arabic}
Bitcoin is a novel monetary system which aspires to replace traditional ones, or at least supplement them. It emphasizes principles such as transparency, efficiency and personal empowerment. Its main technical innovation, the blockchain (responsible for preventing double-spending, \cite{Doublespend}), has received attention in recent months for its applications beyond only a currency.

Inspired by the Austrian school of economics, which dictates that a fixed monetary base is optimal, inflation in Bitcoin is bounded, and the amount of bitcoins in circulation will never exceed 21 million. Inflation consists in ``miners'' performing specific calculations recognized by the network, which gives them a chance to find a block and receive a reward (usually facilitated by a pool, \cite{pools}). The reward is a combination of transaction fees paid by the users whose transactions are confirmed in the block, and a set number of new bitcoins that enter circulation at the miner's hands.

The Bitcoin protocol dictates that the first blocks carry a reward of 50 BTC, and every 210000 blocks (approximately 4 years) the reward is cut in half. At the time of writing this paper, the block reward is 25 BTC, and is expected to be cut to 12.5 BTC soon. This guarantees that, as agreed, there will never be more than 21 million bitcoins ($210000\sum_{i=0}^{\infty}50\cdot2^{-i} = 21\cdot10^6$).

This halving of the reward is known as ``block reward halving'' or simply ``block halving''. These events are considered milestones in the evolution of Bitcoin. By demonstrating how participants in the Bitcoin network, from all around the world and from all walks of life, come together to observe the rules that were set forth in advance and which they all signed up for, they are a testament to the strength and incorruptibility of the Bitcoin protocol.

Because of this, and because block halving is a rare event which only happens once every 4 years, it is widely celebrated in parties thrown all over the world, known as ``block halving parties'' or simply ``block parties''. Some experts have used terms such as ``Awesome'', ``Crazy'' and ``Radical'' to describe block parties (\cite{Promo}), which goes to show the significance of these events.

However, organizing parties of such magnitude requires knowledge of the time of the party; attendees need to clear their schedule, and organizers need to coordinate catering, music, signage, favors, EMS, displays, registration, attractions and other logistics. The statistical nature of block mining, as well as the particulars of the control mechanism that regulates it, make it difficult to predict the exact time of the halving (around which the time of the party should be based).

That said, estimates are possible, and party organizers need as accurate an estimate as possible, not only for the expected halving time, but also for the variance of the prediction (as a function of time left until halving); this can help in determining how long to wait to announce the time of the party.

Some online tools already exist to assist in predicting the halving time (\cite{clock1}, \cite{clock2}, \cite{clock3}), but they do not take all factors into account, and they do not provide the variance of the prediction.

In this paper, We will present several models for making this estimation. In \autoref{sec:fixedhash} we provide the simplest model which assumes a constant hashrate and difficulty. In \autoref{sec:retarget} we take the difficulty retargeting mechanism into account. In \autoref{sec:hashchange} we discuss the effect of a change in network hashrate.

\section{Constant difficulty}\label{sec:fixedhash}
If the Bitcoin network hashrate is constant, and the difficulty is constant and matches the hashrate, a block will be found every 10 minutes on average. Mining blocks consists in trying hashes randomly until one is found that matches the target, which happens with a specific probability. Because of this, the time to find a block is distributed exponentially, with a mean of 10 minutes, and hence (since the exponential distribution has equal mean and standard deviation), a variance of 100 $\min^2$.

If there are $N$ blocks remaining until the halving, the total time until halving (denoted $T$) is a sum of $N$ such random variables. Since expectation is linear, $\mathbb{E}[T]=10\min\cdot N$. The variables are all independent, and hence the variance of their sum is the sum of variances: $\mathbb{V}[T]=100\min^2\cdot N$. This means that the standard deviation is:
\[\sigma = \sqrt{\mathbb{V}[T]}=10\min\cdot\sqrt{N}.\].

If $N$ is large enough, the distribution of $T$ closely matches the normal distribution. This means that tables of areas under the standard normal curve can be used to construct intervals within which the halving is expected to occur with given probabilities. Notably, it is known that a normal variable has about a 68.3\% chance to be within 1 standard deviation of the mean, and a 95.5\% chance to be within 2 standard deviations of the mean.

\textbf{Example.} Suppose the current time is June 2, 2016, 23:50, and there are 5476 blocks remaining until the halving. The expected time to halve is:
\[\mathbb{E}[T]=10\min\cdot5476 = 54760\min = 38\mathrm{day} + 40\min,\]
which means that the estimated halving time is July 11, 00:30. The standard deviation is:
\[\sigma=10\min\cdot\sqrt{5476} = 740\min = 12\mathrm{hr}+20\min,\]
which means the halving has a 68\% chance to be between July 10, 12:10 and July 11, 12:50; and a 95\% chance to be between July 9, 23:50 and July 12, 1:10.

Since some online calculators already provide the expected time based on the above method, and all that remains is finding the standard deviation, it may be more convenient to calculate it using $\mathbb{E}[T]$ directly, instead of the number of blocks. This is done by the formula:
\[\sigma = \sqrt{10\min\cdot\mathbb{E}[T]}.\]

The following conversion table can be handy:

Hour = 60 minutes, Day = 1440 minutes, Week = 10080 minutes, Month = 43830 minutes, Year = 526000 minutes.

\textbf{Example.} The halving is projected to be in 40 days. In minutes, this is $40\cdot1440 = 57600$. Multiplying by 10 gives $576000$. The square root is 759. This means the standard deviation is 759 minutes, which is 12 hours and 39 minutes.

\section{Constant hashrate, variable difficulty}\label{sec:retarget}
The probability for each hash to be a valid block is regulated by a parameter called the ``difficulty''. Specifically, if the difficulty is $D$, the probability for a hash to be a block is $\frac{1}{2^{32}D}$\footnote{More precisely, it is $\frac{2^{16}-1}{2^{48}D}$. We will ignore this distinction.}. This means that if the network hashrate is $H$, the rate at which blocks are found is $\frac{H}{2^{32}D}$.

Every 2016 blocks, the difficulty adjusts based on the time it took to mine the past 2016 blocks\footnote{Due to an off-by-one error, the calculation actually uses the time it took to mine just the past 2015 blocks (\cite{offbyone}). We will ignore this.}. The time it should take is 2 weeks; if the time it actually took is $t$, the difficulty is recalculated as $D:=D\frac{2 \mathrm{week}}{t}$.

Even if the hashrate is constant, statistical variance in the number of blocks found per time unit will cause a change in difficulty; this throws off the naive model in \autoref{sec:fixedhash}, which assumes a constant difficulty.

To take difficulty retargeting into account, we will consider several consecutive intervals of $k=2016$ blocks, which start at end at retargets - the intervals will be enumerated $0, 1, 2, \ldots, n$. Each interval is associated with a difficulty $D_i$, the time $s_i$ that finding those $k$ blocks should have taken on average, the time $t_i$ it actually took to find these blocks, and the ratio $r_i=\frac{t_i}{s_i}$. We will assume we are starting our investigation at the beginning of interval 1, and the halving is $M$ blocks into inerval $n$.

The expected to find a block, given difficulty $D$ and network hashrate $H$, is $\frac{2^{32}D}{H}$. The actual time is distributed exponentially with this mean. The time to find $k$ blocks (which we denoted $t_i$ is the sum of $k$ independent such variables. This has mean $s_i=\frac{2^{32}kD_i}{H}$ and follows the Erlang distribution with rate $\frac{H}{2^{32}D_i}$  and shape parameter $k$, meaning the variance is $k\left(\frac{H}{2^{32}D_i}\right)^2$. The ratio $r_i=\frac{t_i}{s_i}$ therefore follows the Erlang distribution with rate $k$ and shape $k$, which has a mean of 1 and variance of $\frac1{k}$.

By the retarget law, we have $D_{i+1}=\frac{D_i\cdot2\mathrm{week}}{t_i}=\frac{D_i\cdot2\mathrm{week}}{s_ir_i} = \frac{H\cdot2\mathrm{week}}{2^{32}r_i}$, which means that $t_{i+1} = \frac{r_{i+1}D_{i+1}2^{32}}{H}=\frac{r_{i+1}}{r_i}\cdot2\mathrm{week}$. The exception is interval $n$, for which $r_n$ follows $\mathrm{Erlang}(M,M)$, and $t_n=\frac{r_n}{r_{n-1}}M\cdot10\min$.

The total time from start to halving is $T=\sum_{i=1}^nt_i$. The expectation is \[\mathbb{E}[T]=\sum_{i=1}^n\mathbb{E}[t_i]=2\mathrm{week}\cdot\sum_{i=1}^{n-1}\mathbb{E}\left[\frac{r_i}{r_{i-1}}\right]+10\min\cdot M\mathbb{E}\left[\frac{r_n}{r_{n-1}}\right]\]

Since the $r_i$'s are independent, this is equal to
\[2\mathrm{week}\cdot\sum_{i=1}^{n-1}\mathbb{E}[r_i]\mathbb{E}\left[\frac{1}{r_{i-1}}\right]+10\min\cdot M\mathbb{E}[r_n]\mathbb{E}\left[\frac{1}{r_{n-1}}\right]\]
By the properties of the Erlang distribution, $\mathbb{E}[\frac1{r_{i-1}}]=\frac{k}{k-1}=\frac{2016}{2015}$. So we have:
\[\mathbb{E}[T]=\frac{2016}{2015}(n-1)\cdot2\mathrm{week}+\frac{2016}{2015}M\cdot10\min.\]
Note that every additional retarget period increases the time by slightly more than 2 weeks. This means that Bitcoin's schedule is actually $0.05\%$ slower than intended (on top of any slowdown caused by implementation errors, which were ignored in the analysis). Although the effect is minor, this result is novel to the best of the author's knowledge.

More interesting than the expectation, which is only slightly higher than anticipated, is the effect of this mechanism on variance. We have:
\[\mathbb{V}[T]=\sum_{i=1}^n\mathbb{V}[t_i]+2\sum_{i=1}^{n-1}\sum_{j=i+1}^n\mathrm{Cov}(t_i,t_j)\]
\[\mathbb{V}[t_i]=(2\mathrm{week})^2\mathbb{V}\left[\frac{r_i}{r_{i-1}}\right]\]
\[\mathbb{V}\left[\frac{r_i}{r_{i-1}}\right]=\mathbb{E}\left[\left(\frac{r_i}{r_{i-1}}\right)^2\right] - \mathbb{E}\left[\frac{r_i}{r_{i-1}}\right]^2 = \mathbb{E}[r_i^2]\mathbb{E}\left[\frac{1}{r_{i-1}^2}\right] - \left(\frac{k}{k-1}\right)^2 = \]
\[=\frac{k^2}{(k-1)(k-2)}\frac{k+1}{k} - \left(\frac{k}{k-1}\right)^2 = \frac{k(2k-1)}{(k-2)(k-1)^2}\]
\[\sum_{i=1}^n\mathbb{V}[t_i] = (2\mathrm{week})^2\cdot\frac{k(2k-1)(n-1)}{(k-2)(k-1)^2}+(10\min)^2M\frac{k^2(k+M-1)}{(k-2)(k-1)^2}\]
Since $t_i$ is independent of $t_j$ for $j\ge i+2$, we have
\[\sum_{i=1}^{n-1}\sum_{j=i+1}^n\mathrm{Cov}(t_i,t_j)=\sum_{i=1}^{n-1}\mathrm{Cov}(t_i,t_{i+1})=\]
\[=(2\mathrm{week})^2\sum_{i=1}^{n-2}\mathrm{Cov}\left(\frac{r_i}{r_{i-1}},\frac{r_{i+1}}{r_i}\right)+ (2\mathrm{week}\cdot10\min M)\mathrm{Cov}\left(\frac{r_{n-1}}{r_{n-2}},\frac{r_{n}}{r_{n-1}}\right)\]
\[\mathrm{Cov}\left(\frac{r_i}{r_{i-1}},\frac{r_{i+1}}{r_i}\right) = \mathbb{E}\left[\frac{r_i}{r_{i-1}}\cdot\frac{r_{i+1}}{r_i}\right]-\mathbb{E}\left[\frac{r_i}{r_{i-1}}\right]\cdot\mathbb{E}\left[\frac{r_{i+1}}{r_i}\right]=\frac{k}{k-1}-\left(\frac{k}{k-1}\right)^2=\frac{-k}{(k+1)^2}\]
The above holds also for $i=n-1$ because $\mathbb{E}[r_n]=1$. Putting this all together gives:
\begin{eqnarray*}
\mathbb{V}[T] &=& (2\mathrm{week})^2 \left(\frac{k(2k-1)(n-1)}{(k-2)(k-1)^2}-\frac{2(n-2)k}{(k+1)^2}\right) -\\
&-& (2\mathrm{week}\cdot10\min)\frac{2k}{(k+1)^2}+(10\min)^2M\frac{k^2(k+M-1)}{(k-2)(k-1)^2}
\end{eqnarray*}
Notably, the derivative of this with respect to $n$ is $(2\mathrm{week})^2\frac{k (3 + k (11 k-10))}{(k-2) (k^2-1)^2}$. Putting $k=2016$ gives $1100\min^2$, which means that adding an interval of 2016 blocks only adds $1100\min^2$ of variance to the estimate. This is tremendously lower than what the model in \autoref{sec:fixedhash} predicts, which is $201600\min^2$ (the difference is by a factor of 200, that is, we have only $0.5\%$ of the marginal variance). The reason is that any change in the luck on one interval creates a correction on the next, which almost exactly cancels out. We are left, basically, with only the variance of the first and last intervals, which is roughly
\[\mathbb{V}[T]=(100\mathrm{min}^2)\left(M+\frac{M^2}{k}+\frac{8133000}{k}\right)\]

Once the variance has been found, the standard deviation is $\sigma=\sqrt{\mathbb{V}[T]}$, which can be used as in the examples in \autoref{sec:fixedhash}.

Notably, the second block halving happens at block 420000, and the prior difficulty retarget is at block 419328. This means that for predicting this particular halving we have $M=420000-419328 =672$. Putting this and $k=2016$ gives $\mathbb{V}[T]=493000\min^2$ and $\sigma = 702\min = 11\mathrm{hr}+42\min$. In other words, at any time more than a full retarget period prior to the halving, the standard deviation of the estimate is about 11 hours and 42 minutes.

\section{Hashrate changes}\label{sec:hashchange}
If the network hashrate changes, the halving time will change accordingly. If the hashrate increases by a small amount $x\%$, then for a period of about one difficulty retargeting interval, blocks will be found $x\%$ faster than normal, after which we will return to difficulty equilibrium. During this time, the schedule will move ahead (that is, the halving will be sooner) by $x\%\cdot(2\mathrm{week})$. For example, if the hashrate increases by 10\%, the halving will arrive sooner by roughly 20\% of a week, which is 33.6 hours. The same effect, in reverse, happens if the hashrate decreases. (Note that a slightly different calculation is needed if the change is larger).

A continuous increase of hashrate, spanned over a long enough period well before the halving, can be considered as the compounding of multiple small increments. If the old hashrate is $H_1$ and the new hashrate is $H_2$, the schedule change is $\log\left(\frac{H_2}{H_1}\right)(2\mathrm{week})$. This works for an increase of any size (including by orders of magnitude), as long as it is spread over a long period.

\textbf{Example.} The model of the previous sections predicts that the halving will happen on July 11 at 01:00. However, the hashrate is set to increase by a total of 50\%, over a prolonged period of time which is well before the halving. Then the schedule will shift by $\log1.5\cdot(2\mathrm{week}) = 0.405\cdot(2\mathrm{week}) = 5\mathrm{day}+16\mathrm{hr}$, which means that the halving will happen on July 6, 9:00.

If the hashrate change happens near the halving - not a full retarget interval apart - its effect will be roughly proportional to the number of blocks remaining. For example, if the hashrate increase by 5\%, 300 blocks before the halving, the schedule will move ahead by $5\%\cdot300=15$ blocks, which is 150 minutes, or 2 and a half hours.

Because the advent of new generations of mining hardware can significantly alter the hashrate, this effect poses much greater uncertainty than the statistical variance analyzed in previous sections. However, if an organizer has information about the hashrate trajectory, he can use the principles described in this section to incorporate it into the prediction.

\section{Conclusions}
In this paper we have discussed the importance of block parties, outlined some known techniques to predict their timing, and also developed a novel model which takes into account the effect of long-term difficulty retargeting on variance.

\newpage
\bibliographystyle{plain} 
\bibliography{BlockParty}
\end{document}